\documentclass{article}
\usepackage{LaThuileFPSpro2}
\begin{document}
%

%
\title{ 
  $W$ Boson Production and Mass at the Tevatron
  }
\author{
  Oliver Stelzer-Chilton      \\
  (on behalf of the CDF and D{\O} Collaborations) \\
  {\em University of Toronto, Ontario, Canada} \\
  {\em Email: oliversc@fnal.gov} \\
  }
\maketitle

\baselineskip=11.6pt

\begin{abstract}
  The CDF and D{\O} collaborations have analyzed up to $\sim$200 pb$^{-1}$ of Run 2 physics data
  to measure $W$ production properties such as the $W$ cross section, the $W$ width,
  lepton universality and the $W$ charge asymmetry. From the cross section measurements,
  CDF obtains a lepton universality of $\frac{g_{\mu}}{g_{e}} = 0.998\pm0.012$
  and $\frac{g_{\tau}}{g_{e}} = 0.99\pm0.04$ and an indirect $W$ width of
  $\Gamma_W$=2079$\pm$41 MeV. D{\O} measured the $W$ width directly and finds
  $\Gamma_W$=2011$\pm$142 MeV. CDF has estimated the uncertainties 
  on the $W$ boson mass measurements in the electron and muon decay channels and obtains
  an overall uncertainty of 76 MeV. 
\end{abstract}
%
%
\section{Introduction}
The properties of the $Z$ boson have been measured to very high precision
at LEP\cite{zpdg}. Naturally one wants to match this precision for the charged
carriers of the electroweak interaction. Over the next few years the
Tevatron is the only accelerator which can produce $W$ bosons.
Measuring the properties of the $W$ boson to a very high precision is an important
test of the Standard Model. From the
measured $W$ cross section, one can infer an indirect measurement of the
$W$ width and lepton universality. Since at the Tevatron the $W$ bosons are 
produced through quark anti-quark annihilation, a significant uncertainty 
for all direct electroweak measurements comes from the knowledge of the
parton distributions inside the proton. The probability of finding a parton 
carrying a momentum fraction $x$ within the incoming proton is expressed in the 
parton distribution function (PDF). The measurement of the $W$ charge
asymmetry provides important input on the ratio of the $u$ and $d$ quark components
of the PDF and will help to further constrain parton distribution functions. 
The $W$ boson mass serves
as a test of the Standard Model, but through radiative corrections is also
sensitive to hypothetical new particles. Together with a
precise measurement of the top quark mass\cite{top}, the $W$ boson mass constrains the
mass of the Higgs boson, which has not yet been observed experimentally.

Both CDF and D{\O} are multi-purpose detectors. They consist of tracking 
systems surrounded by calorimeter and muon identification systems. 
CDF's tracking system consists of a wire drift chamber (the Central Outer Tracker) and a 7-layer silicon
microstrip vertex detector (SVXII) immersed in a 1.4 T magnetic field. A lead (iron) 
scintillator sampling calorimeter is used for measuring electromagnetic 
(hadronic) showers. D{\O} employs a silicon microstrip tracker (SMT) and a central
fiber tracker (CFT), both located in a 2 T magnetic field. The sampling calorimeter
consists of liquid argon and uranium.

Since the hadronic decay of the $W$ boson has an extremely large background
originating from strongly interacting processes, CDF and D{\O} use the clean leptonic
decays to study the $W$ boson. The signature is a high energy lepton 
with large missing transverse momentum originating from the neutrino, 
which does not interact with the detector. The momentum balance in the
direction of the beam is unconstrained and as a result, the $W$ events
are studied in the plane transverse to the beam. A typically used quantity is the
transverse mass:
\begin{equation}
  M_T = \sqrt{2 p_T^l p_T^{\nu}(1-cos(\Delta\phi))},
  \label{trmass} 
\end{equation}
which is similar to the invariant mass, just in the two transverse dimensions.
If not otherwise stated, we restrict the lepton identification to the well
instrumented central region of $|\eta|<1$. 

$Z$ boson events are identified by two high energy leptons. 
These events have very low background.
\section{Inclusive $p\bar p\rightarrow W/Z+X$ Cross Section Measurements}
\begin{figure}[h]
  \vspace{9.0cm}
  \includegraphics{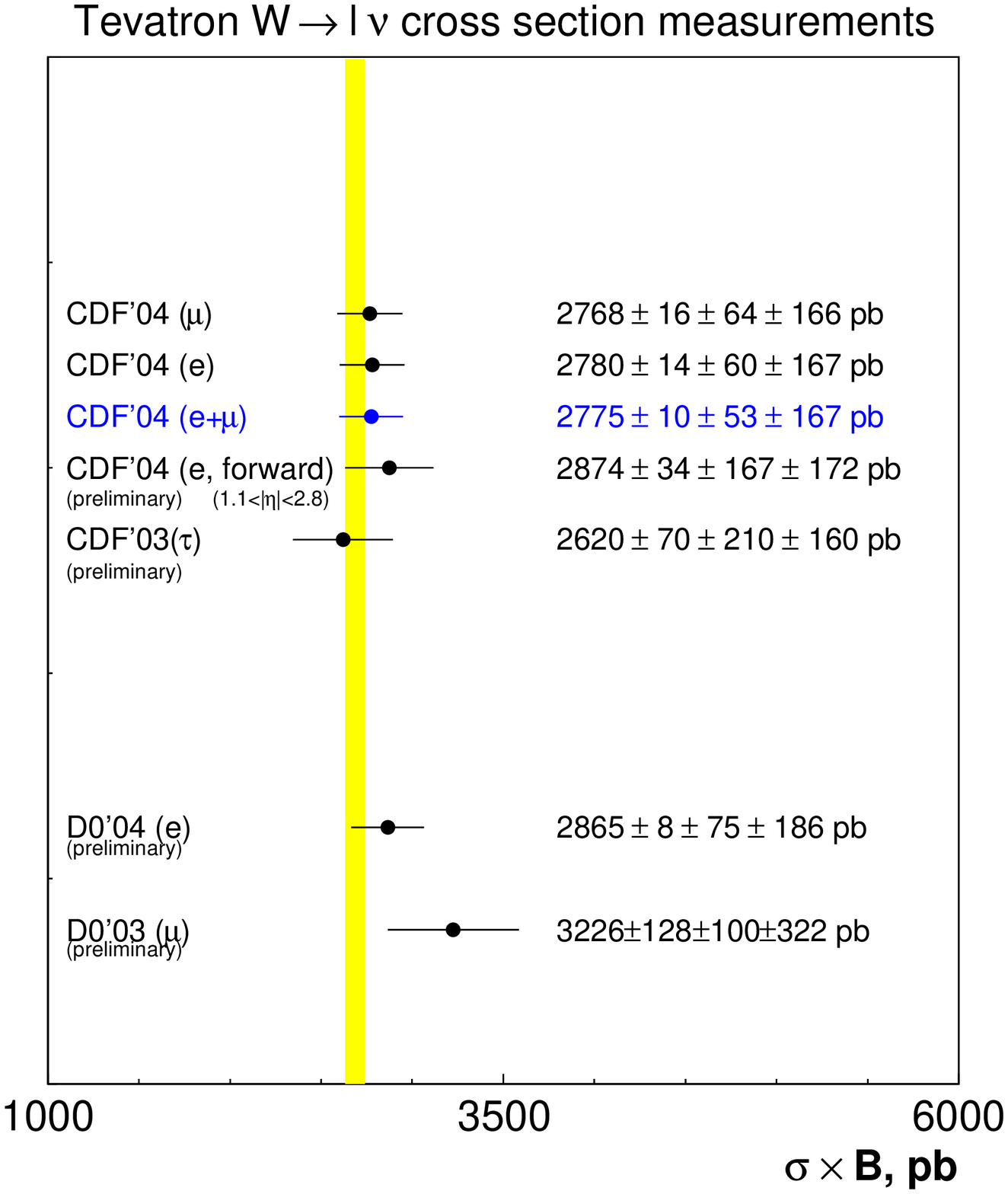}
  \includegraphics{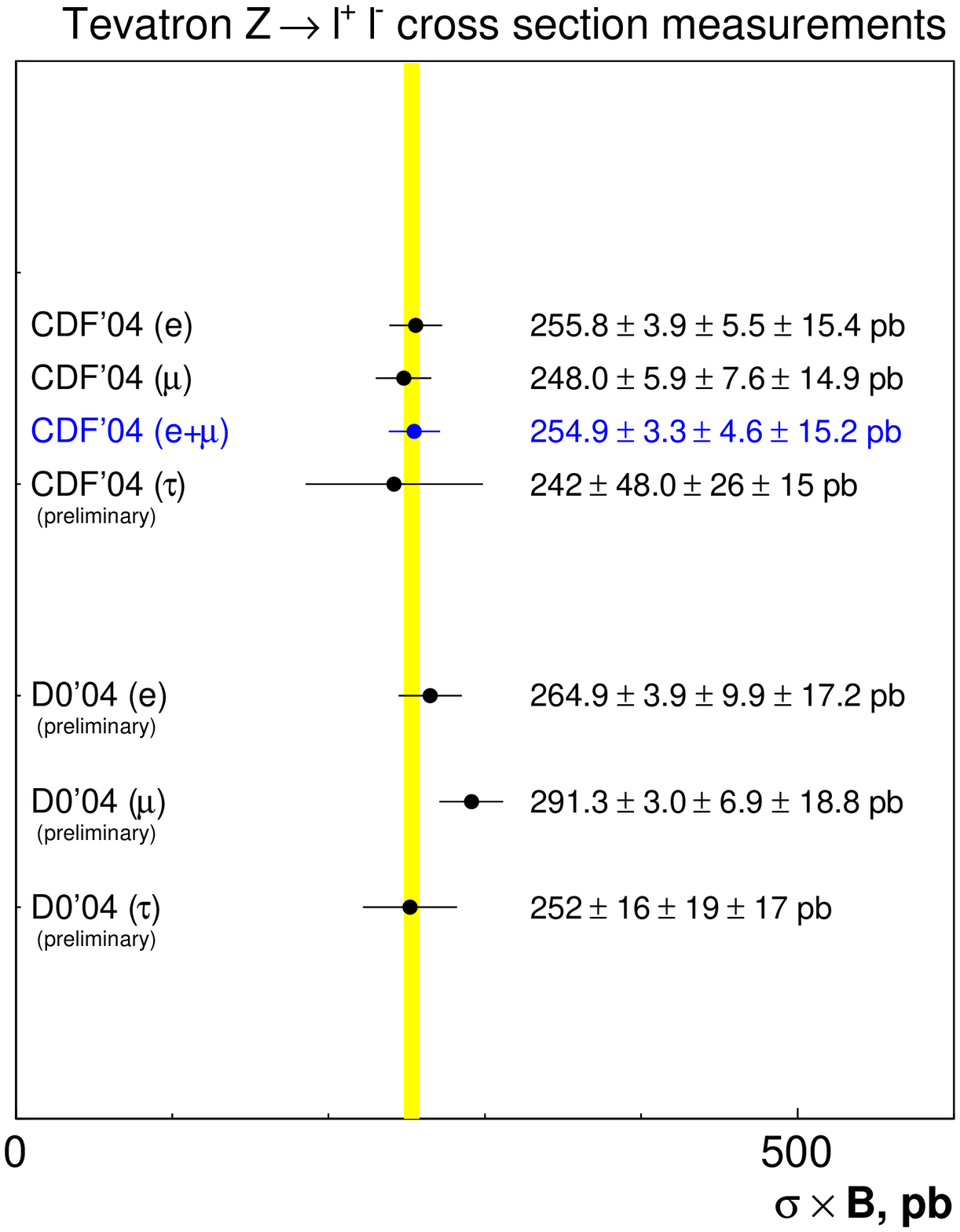}    
  \caption{\it
    Summary of various CDF and D{\O} $W$ and $Z$ cross section measurements in all three leptonic decay channels, using different datasets and sub-detectors.
    The uncertainties are listed in the following order: statistical, systematic, and luminosity.
    \label{WZX} }
\end{figure}

$W$ and $Z$ bosons are identified by their leptonic decays to electrons, muons and taus, 
from which the total rates $\sigma\times$Br($W\rightarrow l\nu$) and 
$\sigma\times$Br($Z\rightarrow l l$) are obtained. The cross section times branching
ratio is calculated as follows:
\begin{equation}
  \sigma\times Br(p\bar p\rightarrow W/Z\rightarrow ll) = \frac{N_{cand}-N_{bkg}}{A\epsilon L}.
  \label{xseceq} 
\end{equation}
The $W$ and $Z$ boson cross sections have been measured by CDF\cite{wzcross} with different datasets
in different sub-detectors. Figure \ref{WZX} 
shows a summary of the CDF and D{\O} cross section measurements in all leptonic decay modes.
All measurements show good agreement with NNLO calculations\cite{nnlo}, represented by the vertical
band.
\subsection{Lepton Universality in $W$ Decays}
Lepton universality in $W$ decays can be tested by extracting the ratio of the
electroweak couplings $g_{\mu}/g_e$ and $g_{\tau}/g_e$ from the measured ratio of
$W\rightarrow l\nu$ cross sections. The $W\rightarrow l\nu$ couplings are related
to the measured production cross section ratio $U$ as follows:
\begin{equation}
  U=\frac{\sigma\times Br(W\rightarrow l\nu)}{\sigma\times Br(W\rightarrow e\nu)}=\frac{\Gamma(W\rightarrow l\nu)}{\Gamma(W\rightarrow e\nu)} = \frac{g_{l}^2}{g_{e}^2}
  \label{lepuni} 
\end{equation}
In this ratio, important systematic uncertainties cancel. The results obtained are\cite{wzcross}:
\begin{equation}
  \frac{g_{\mu}}{g_{e}} = 0.998\pm0.012
  \label{lepunimu} 
\end{equation}
\begin{equation}
  \frac{g_{\tau}}{g_{e}} = 0.99\pm0.04
  \label{lepunitau} 
\end{equation}
where the largest systematic uncertainty comes from event selection efficiencies. Since these
efficiencies are measured using the $Z\rightarrow l l$ sample, the uncertainty will decrease
as more $Z$ bosons are collected.

\subsection{Indirect $W$ Width Determination}
The ratio $R$ of the cross section measurements for $W$ and $Z$ bosons can be used to extract
the total width of the $W$ boson. $R$ can be expressed as:
\begin{equation}
   R=\frac{\sigma(p\bar p\rightarrow W)}{\sigma(p\bar p\rightarrow Z)}\frac{\Gamma(W\rightarrow l\nu)}{\Gamma(Z\rightarrow l l)}\frac{\Gamma(Z)}{\Gamma(W)}.
  \label{indwidth} 
\end{equation}
Using the very precise measurement of $\Gamma(Z\rightarrow l l)$/$\Gamma(Z)$ from LEP and NNLO calculations
of $\sigma(p\bar p\rightarrow W)/\sigma(p\bar p\rightarrow Z)$, together with the Standard Model prediction
of $\Gamma(W\rightarrow l \nu)$ one can extract $\Gamma(W)$ from equation \ref{indwidth}.
\begin{table}[h]
  \centering
  \caption{ \it Summary of indirect $W$ width measurements.
    }
  \vskip 0.1 in
  \begin{tabular}{|l|c|c|} \hline
    Channel & $\Gamma(W) (MeV) $ & L ($pb^{-1}$) \\
    \hline
    $W\rightarrow e \nu + W\rightarrow \mu \nu$   & 2079 $\pm$ 41 &  72                    \\
    $W\rightarrow \mu \nu$  & 2056 $\pm$ 44  & 194                 \\
    World average & 2124 $\pm$ 41 & \\
    \hline
  \end{tabular}
  \label{indwidthsum}
\end{table}
Table \ref{indwidthsum} shows the values from CDF for two different datasets, 
together with the current world average (not including these measurements). 
The indirect width measurements show good agreement and have competitive uncertainties.

\section{Direct $W$ Width Measurement}
D{\O} has measured the $W$ boson width directly in the electron decay channel\cite{d0width}. The measurement uses an 
integrated luminosity of 177 pb$^{-1}$. 
The width is determined by normalizing the predicted signal and background transverse
mass distribution in the region of 50 GeV$<$$M_T$$<$100 GeV and then fitting the predicted shape to
the candidate events in the tail region 100 GeV$<$$M_T$$<$200 GeV,
\begin{figure}[h]
  \vspace{8cm}
  \includegraphics{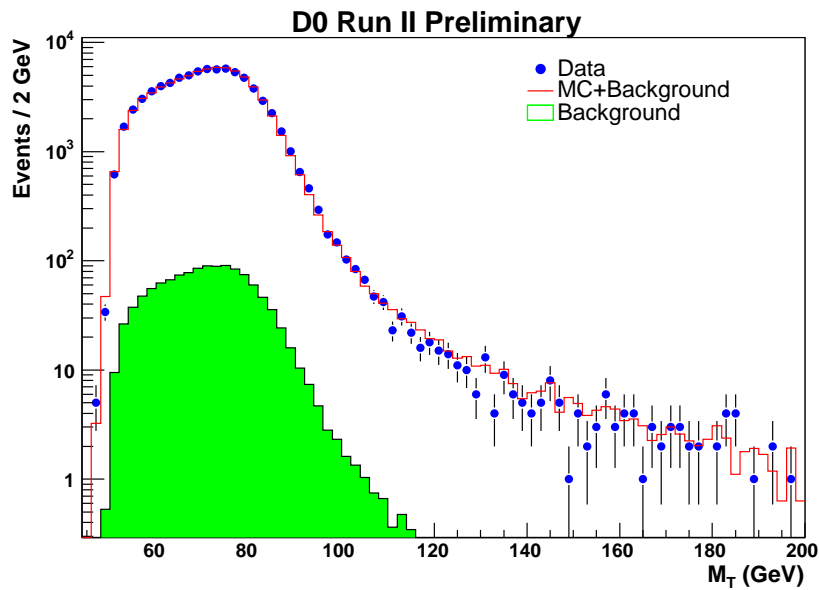}
  \caption{\it
    $W$ width extraction from high transverse mass tail.
    \label{d0width} }
\end{figure}
which is most sensitive to the width.
Figure \ref{d0width} shows the transverse mass distribution. The measurement yields
$\Gamma_W=$2011$\pm$93(stat)$\pm$107(syst) MeV, which is in good agreement with the
world average, an improvement over the D{\O} Run 1 measurement\cite{d0widthrun1}, and competitive to the
CDF Run 1 measurements in the muon and electron decay channels\cite{wcombo}.

\section{$W$ Charge Asymmetry}
The $W$ bosons at the Tevatron are produced predominantly through annihilation of valence $u$ ($d$)
and anti-$d$ (anti-$u$) quarks inside the proton and anti-protons for $W^+$ ($W^-$) production.
Since $u$ quarks carry, on average, a higher fraction of the proton momentum than $d$ quarks, a $W^+$ 
tends to be boosted in the proton direction, while a $W^-$ is boosted in the anti-proton direction. 
This results in a non-zero forward-backward charge asymmetry, defined as:
\begin{equation}
A(y_W)=\frac{d\sigma(W^+)/dy_W-d\sigma(W^-)/dy_W}{d\sigma(W^+)/dy_W+d\sigma(W^-)/dy_W},
  \label{etaasym} 
\end{equation}
where $y_W$ is the rapidity of the $W$ bosons and $d\sigma(W^{\pm})/dy_W$ is the differential cross section
for $W^+$ or $W^-$ boson production. However, because the $p_Z$ of the neutrino is unmeasured, $y_W$ cannot
be directly determined, and we instead measure:
\begin{equation}
A(\eta_e)=\frac{d\sigma(e^+)/d\eta_e-d\sigma(e^-)/d\eta_e}{d\sigma(e^+)/d\eta_e+d\sigma(e^-)/d\eta_e},
  \label{etaasym} 
\end{equation}
where $\eta_e$ is the electron pseudorapidity. Therefore, the observed asymmetry is a convolution
of the aforementioned charge asymmetry and the Standard Model $V-A$ couplings describing the $W\rightarrow e\nu$
decays. 
A measurement of the forward-backward charge asymmetry is sensitive to the ratio of the $u$ and $d$ 
quark components of parton distribution functions.
CDF has measured this asymmetry in the electron channel up to a pseudorapidity of $|\eta|$$<$2.5 
using 170 pb$^{-1}$\cite{wcharge}.
\begin{figure}[h]
  \vspace{8cm}
  \includegraphics{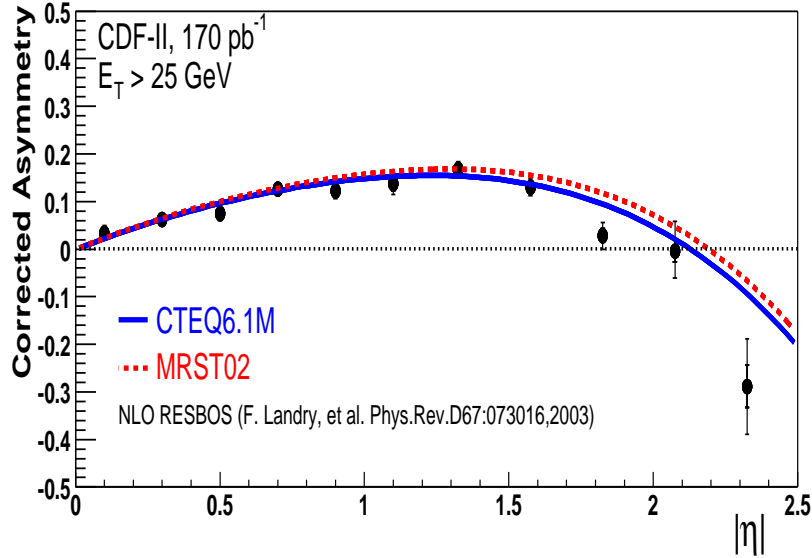}
  \caption{\it
    Measured asymmetry corrected for the effects of charge misidentification and background contributions.
    \label{Wcharge} }
\end{figure}
Figure \ref{Wcharge} shows the measured asymmetry corrected for the effects of 
charge misidentification and background contributions. The predictions using the latest 
CTEQ and MRST PDFs are overlaid. This measurement will provide important input for the next generation 
of PDFs.

\section{$W$ Mass}
Since its discovery in 1983\cite{wdisc}\cite{wdisc2}, the $W$ boson mass has been measured with increasing precision.
From an initial uncertainty of 5 GeV, the uncertainty of the $W$ mass has been reduced to 42 MeV
from the LEP experiments\cite{wlep} and to 59 MeV from the Tevatron experiments\cite{wcombo}. CDF has analyzed the first
200 pb$^{-1}$ of Run 2 data and estimated the corresponding $W$ boson mass uncertainty in the electron
and muon decay channels. The uncertainty includes contributions from statistics, production and decay 
modeling, lepton energy scale and resolution, hadronic recoil and resolution, and backgrounds.

There are two important aspects to a precision $W$ mass measurement: Calibration
of the detector to the highest possible precision, and simulation of 
the transverse mass spectrum, which cannot be predicted analytically. The simulation
includes the production modeling and detector effects and
produces transverse mass templates for a range of $W$ boson masses. Since backgrounds
contaminate the signal, they are included in the templates. The $W$ mass is 
extracted from a maximum likelihood fit to the transverse mass spectrum.

\subsection{Production Model}
The uncertainty in the modeling of the $W$ boson production and decay results from parton
distribution functions, QED radiative corrections, the transverse momentum of the $W$ boson and the
$W$ boson width. 

The parton distribution functions affect the $W$ mass through the limited acceptance of the 
detector for the $W$ decay lepton. The uncertainty has been determined using the set of 40 CTEQ6
PDFs\cite{cteq}, which explore the uncertainty on the 20 orthogonal eigenvector directions in 
parameter space. Each eigenvector direction corresponds to some linear combination of
PDF parameters. The resulting uncertainty is $\Delta M_W(e,\mu)=\pm$15 MeV. A cross check using
the latest MRST\cite{mrst} PDF falls within this estimate.

The dominant higher-order QED effect on the $W$ boson mass is photon radiation off the final-state
charged lepton. Additional QED uncertainties arise from multi-photon radiation, initial state
radiation and radiation from interference terms, none of which are included in the simulation
used to extract the $W$ boson mass. The uncertainty from QED corrections is $\Delta M_W(\mu)=\pm$
20 MeV in the muon channel and $\Delta M_W(e)=\pm15$ MeV in the electron channel.

The initial-state QCD radiation in vector boson production is constrained by a phenomenological
parametrization of the $Z$ boson $p_T$ measurement from the previous collider run. The parameters 
are used for the modeling of the $W$ $p_T$ distribution and their uncertainties result in $\Delta
M_W(e,\mu)=\pm$13 MeV.

The uncertainty on the $W$ boson width affects the falling Jacobian edge and leads to $\Delta
M_W(e,\mu)=\pm$12 MeV.
\subsection{Lepton Momentum/Energy Scale and Resolution}
The lepton momentum measurement is based fundamentally on the calibration of the tracking
wire chamber (COT). After the calibration of the track momentum and resolution using the
muon decays of precisely known resonances, the energy scale of the electromagnetic calorimeter
is calibrated using the ratio of calorimeter energy to track momentum ($E/p$) of electrons.

\begin{figure}[h]
  \vspace{8cm}
  \includegraphics{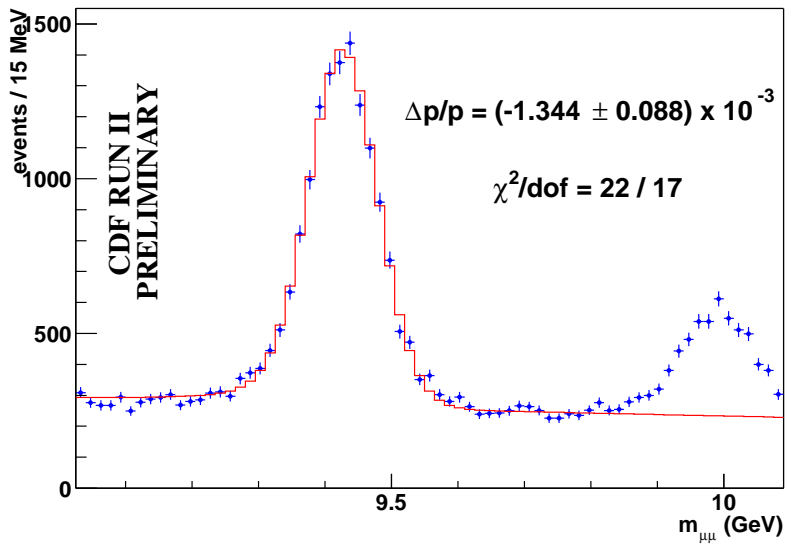}
  \caption{\it
    The reconstructed invariant mass of muon candidate pairs in the $\Upsilon(1S)$ region.
    The fractional difference between the measured and PDG mass is shown.
    \label{ups} }
\end{figure}

The quarkonium resonance decays $J/\Psi\rightarrow\mu\mu$ and $\Upsilon(1S)\rightarrow\mu\mu$ 
are used to set the momentum scale (Figure \ref{ups}).
The passive material in the simulation is tuned such that the reconstructed $J/\Psi$ mass
is constant as a function of mean track curvature. The measured momentum scale is the mean 
of the individual $J/\Psi$ and
$\Upsilon(1S)$ scales. The systematic uncertainty is taken as half the difference between
the extracted scales which results in $\Delta M_W(e,\mu)=\pm$15 MeV.

The track resolution is parametrized in the simulation by the individual hit resolution
and by the hit multiplicity on the track. Muons from decays of $Z$ bosons are used to determine
the resolution. The resulting uncertainty corresponds to $\Delta M_W(e,\mu)=\pm$12 MeV.
An additional uncertainty of $\Delta M_W(e,\mu)=\pm$20 MeV is assigned for tracking chamber
misalignments.

\begin{figure}[h]
  \vspace{8cm}
  \includegraphics{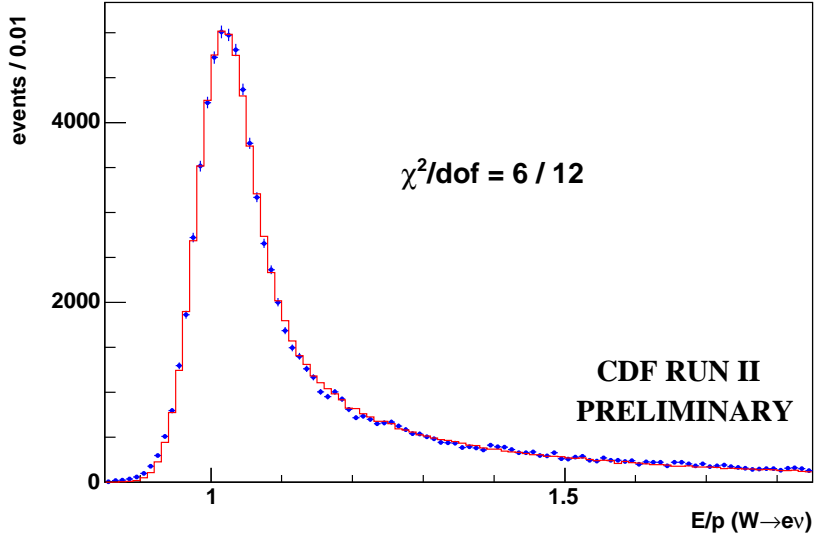}
  \caption{\it
    The $E/p$ distribution of electrons from $W$ boson decays. The region between 0.95 and 1.1
    is used for the calibration of the electromagnetic energy scale.
    \label{eop} }
\end{figure}

The $E/p$ distribution of electrons from $W$ boson decays is used to calibrate the electromagnetic
energy scale of the calorimeter (Figure \ref{eop}). 
The statistical uncertainty and the uncertainty from the
momentum scale results in $\Delta M_W(e)=\pm$35 MeV.
Additional energy scale uncertainties arise
from the calibration of the detector passive material and from the calorimeter non-linearity.
The passive material was measured during detector construction and is tuned using electrons
from photon conversions. A final tuning uses the tail of the $E/p$ distribution which is
sensitive to the amount of material modeled in the simulation. The uncertainty
on the passive material results in $\Delta M_W(e)=\pm$55 MeV. The calorimeter non-linearity
is determined from the $E_T$ dependence of the energy scale. After applying a correction, the
uncertainty on the slope results in $\Delta M_W(e)=\pm$25 MeV.

The calorimeter resolution is parametrized as $\sigma E_T/E_T$=13.5\%/$\sqrt E_T$$\oplus\kappa$,
where $\kappa$ is determined from the width of the $E/p$ signal. The uncertainty on $\kappa$
results in $\Delta M_W(e)=\pm$7 MeV.

The $Z$ boson masses are used as cross checks for the momentum and energy scales.
\subsection{Recoil Scale and Resolution}
The hadronic recoil is measured by summing over the energy in all calorimeter
towers, excluding the lepton towers. The simulation removes an equivalent set of 
towers by subtracting the mean underlying event energy as measured from adjacent towers. 
The uncertainty on the measurement of this underlying event energy
results in $\Delta M_W(e)=\pm$15 MeV and $\Delta M_W(\mu)=\pm$10 MeV uncertainties on the $W$ mass. 

The hadronic recoil scale is the ratio of measured to true recoil. It is parametrized as a function
of the true recoil and tuned using $Z$ events where both decay leptons
are reconstructed and the $Z$ boson $p_T$ can be reconstructed precisely from the lepton
momentum measurements. The uncertainty on the parametrization results in $\Delta M_W(e,\mu)=\pm$20 MeV.

The recoil resolution model incorporates terms from the underlying event, which are modeled
with generic inelastic collisions, and from hadronic jet resolution. The resolution
uncertainty results in $\Delta M_W(e,\mu)=\pm$42 MeV.
\subsection{Backgrounds}
The backgrounds in the $W$ boson data sample include $W\rightarrow\tau\nu$, $Z\rightarrow ll$
where one lepton is outside the detector acceptance and not reconstructed, hadronic jets, 
where one jet mimics a lepton, cosmic rays, where one leg of the cosmic track is not reconstructed, 
and kaon decays, where the kaon track is misreconstructed, resulting in large apparent muon momentum. 
The background measurement uncertainties result in $\Delta M_W(e,\mu)=\pm$20 MeV.
\subsection{Mass Fits and Total Uncertainty}
\begin{figure}[h]
  \vspace{8cm}
  \includegraphics{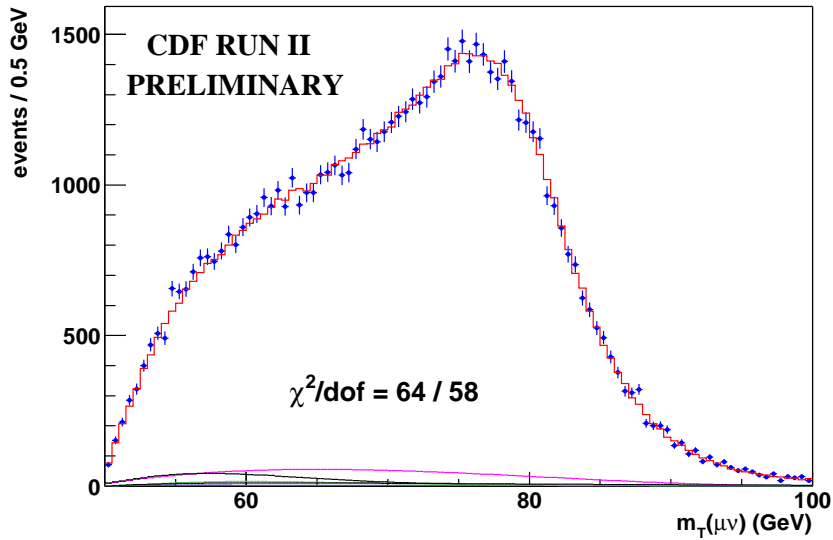}
  \caption{\it
    The $M_T$ distribution in $W$ boson decays to muons. The points represent the data, the
    histogram the simulation with backgrounds added. The region between 60-90 GeV is 
    used to fit the $W$ boson mass.
    \label{mt} }
\end{figure}

After including all measurement components, CDF obtains transverse mass (Figure \ref{mt})
and transverse energy distributions which are well modeled. Table \ref{syst}
summarizes the uncertainties for the $M_T$ fits in the electron and muon decay channels.
\begin{table}[h]
  \centering
  \caption{ \it The uncertainty on the $W$ boson mass measurement using $\sim$200 pb$^{-1}$ of
  Run 2 CDF data. The CDF Run 1b uncertainties are shown for comparison.
    }
  \vskip 0.1 in
  \begin{tabular}{|l|c|c|} \hline
    Uncertainty & Electrons (Run 1b) & Muons (Run 1b) \\
    \hline
    Production and Decay Model  & 30 (30) &  30 (30)                    \\
    Lepton Energy Scale and Resolution & 70 (80)  & 30 (87)                 \\
    Recoil Scale and Resolution & 50 (37) & 50 (35) \\
    Backgrounds & 20 (5) & 20 (25)\\
    Statistics & 45 (65) & 50 (100) \\
    \hline
    Total & 105 (110) & 85 (140) \\
    \hline
  \end{tabular}
  \label{syst}
\end{table}
For comparison the uncertainties from the previous collider run\cite{wmassrun1b} (Run 1b) are also included. 
The overall uncertainty is 76 MeV.
The $W$ boson mass fit results are currently blinded with a constant offset. The offset
will be removed when further cross checks have been completed.
\section{Summary}
The $W$ boson physics program at the Tevatron is very successful. CDF and D{\O} have measured
the inclusive $W$ and $Z$ cross sections in all three leptonic decay channels, which show good
agreement with NNLO calculations. From the cross section measurements, CDF has extracted
competitive measurements on lepton universality and an indirect measurement of the $W$ boson width.
D{\O} has measured the $W$ boson width directly in the electron channel with an uncertainty
smaller than the Run 1 value. The new CDF $W$ charge asymmetry will help to further constrain 
the uncertainties of parton distribution functions, which affect all the aforementioned measurements.
With the addition of 600 pb$^{-1}$ of data on tape, these measurements will further constrain the
Standard Model.

CDF has determined the uncertainty on the $W$ boson mass with the first $\sim$200 pb$^{-1}$ of Run 2 data
to be 76 MeV, which is lower than its Run 1 uncertainty of 79 MeV. With the additional data
to come, Run 2 promises the world's highest precision measurement of the $W$ boson mass, with an
anticipated uncertainty of 30 MeV for 2 fb$^{-1}$.
\section{Acknowledgments}
I would like to thank my colleagues from the CDF and D{\O} electroweak groups for their hard
work and input to this talk.

\end{document}